\documentclass[final]{ws-mpla_arxiv}

\newcommand{\lsim}{\;\raisebox{-0.9ex}{$\textstyle\stackrel{\textstyle<}
           {\sim}$}\;}
\newcommand{\gsim}{\;\raisebox{-0.9ex}{$\textstyle\stackrel{\textstyle>}
           {\sim}$}\;}

\begin{document}

\markboth{Are R.~Raklev}{Massive Metastable Charged (S)Particles at the LHC}

\catchline{}{}{}{}{}

\title{MASSIVE METASTABLE CHARGED (S)PARTICLES AT THE LHC}

\author{\footnotesize ARE R.~RAKLEV}

\address{Department of Applied Mathematics and Theoretical Physics,
University of Cambridge, Wilberforce Road, Cambridge CB3 0WA, UK\\
A.Raklev@damtp.cam.ac.uk}

\maketitle


\begin{abstract}

This brief review deals with recent interest in the prospects of
observing a Massive Metastable Charged Particle (MMCP) at the Large
Hadron Collider (LHC), and measuring its properties there. We discuss
the motivation for scenarios with MMCPs in a phenomenological context,
focusing on supersymmetric models that allow us to explore the
expected experimental signatures of MMCPs at the LHC. We review
current bounds and give estimates of the LHC reach in terms of MMCP
masses.

\keywords{Review; metastable; MMCP; supersymmetry; LHC}
\end{abstract}

\ccode{PACS Nos.: 12.60.Jv, 13.85.Rm, 14.80.Ly}

\section{Introduction}	

With the turn-on of the LHC, the High Energy Physics community will
have access to energies that have hitherto been un-probed in lab-based
experiments. It is then perhaps natural that at this time there is a
plethora of ideas for what New Physics might hide at these
energies. Three main ideas seem to underpin most of the work currently
being done: i) the need for a viable Dark Matter candidate, ii) the
want of an explanation of the hierarchy problem in the Standard Model,
and iii) the desire for a Grand Unified Theory (GUT).

Solutions to any one of these problems seem to require an extension of
the particle content of the Standard Model, and such extensions often
predict a rich New Physics sector at the TeV scale. This is not
surprising, since both the Weakly Interacting Massive Particle (WIMP)
interpretation of Dark Matter and the scale of electro-weak symmetry
breaking point to this energy range, if not even lower.

In exploring the TeV scale with the LHC, most suggested New Physics
searches focus on the decay products of short-lived resonances. Weakly
interacting Dark Matter candidates, assumed stable, appear only as
missing energy signatures. However, several mechanisms are known that
could result in Massive Metastable Charged Particles
(MMCPs),
\footnote{By metastable we mean any particle that is stable on
the scale of a lab-based detector experiment, {\it i.e.} that normally
passes through the detector before decaying. The average distance
travelled by an MMCP with energy $E$ and mass $m$ before it decays is
given by $\gamma c\tau$, where $\gamma=E/m$. Charged particles are in
this context electromagnetically charged and/or have strong
interactions.}
with a broad range of possible lifetimes. The long
lifetime for these particles is often intimately connected with the
stability of a Dark Matter candidate. The spectacular signatures of
such models and their great physics potential has attracted a lot of
attention over the last years. Indeed, the recent popularity of MMCPs
is perhaps best gauged by the multitude of names they have been given
in the literature: Stable Massive Particles (SMPs), Long-Lived Charged
Massive Particles (CHAMPs), Heavy Stable Charged Particles (HSCPs),
Charged Massive Stable Particles (CMSPs), and so on.

To focus our discussion within the limited space available in this
brief review we will use supersymmetric models to illustrate the most
important properties of MMCPs at the LHC. There are certainly many
interesting possibilities beyond supersymmetry; a more complete
listing of MMCPs candidates, that demonstrate their seemingly
ubiquitous nature in New Physics models, can be found in Section~2 of
the excellent review by Fairbairn {\it et al.}~given as
Ref.~\refcite{Fairbairn:2006gg}. Nevertheless, in terms of the
expected LHC phenomenology, most other candidates are very similar to
one or more supersymmetric MMCPs, with the exception of MMCPs with
magnetic charge or fractional electric charges. We will also refrain
from any extensive discussion of the cosmological implications of
MMCPs, although these do contribute in setting bounds on MMCP
properties.

We shall begin this review by looking briefly at possible mechanisms
behind the stability of massive particles, exemplified by
supersymmetric models that give MMCPs, in Section~\ref{sec:models},
before we discuss the experimental signatures and issues at colliders
in Section~\ref{sec:exp}. We then review the current bounds on
sparticle MMCPs in Section~\ref{sec:bounds} and give our expectations
for the LHC in terms of the discovery reach and the measurement of
MMCP properties in Section~\ref{sec:LHC}, before we conclude.

\section{Massive Metastable Charged Particles in Supersymmetry}
\label{sec:models}

In the Standard Model (SM) we have a wide selection of different types
of long-lived particles: the stable elementary --- or so we believe
--- electron, the metastable elementary muon, and (very) long-lived
composite particles such as neutrons and protons. These SM particles
illustrate very well various possibilities for stability mechanisms:
\begin{romanlist}[(iii)]
\item Lightest state carrying a conserved quantum number (electron, proton).
\item Suppressed (effective) coupling (muon).
\item Lack of phase space for decay (neutron).
\end{romanlist}
The electron has no lighter electrically charged state in the SM,
while the proton, although decays such as $p\to\pi^0e^+$ are
kinematically allowed, is protected by the conservation of lepton- and
baryon-number. It is interesting to note the different origins of
these conservation laws; charge conservation is the result of a local
gauge symmetry, while lepton- and baryon-number are global non-gauge
accidental symmetries of the SM that can be broken by higher
dimensional operators. Clearly both mechanisms can be used to provide
MMCPs in New Physics scenarios.

In the second mechanism metastability comes about through small
couplings to kinematically allowed decay products. Very small
couplings on their own raises the question of fine tuning in the
theory. However, an effective coupling can be small due to large
differences in scales between the decaying particle, and some mediator
of the decay. This is the case for the muon whose decay proceeds
through an effective four-fermion dimension-six operator with coupling
$g^2m_\mu^2/8m_W^2$.

Another option is the lack of phase space for an otherwise allowed
decay. Again mass degeneracies may seem fine tuned for elementary
particles, less so for composite particles that get their mass from
the same dynamics. In the SM the neutron decays as $n\to
p^+e^-\bar\nu_e$, where the available kinetic energy of $0.8$~MeV to
the final state is tiny compared to the neutron mass, giving a very
long lifetime. The mass degeneracy between the proton and the neutron
is a fortunate consequence of QCD dynamics. Stability mechanisms that
are not seen in the SM particle spectrum, have been
suggested. Topological defects such as magnetic monopoles and Q-balls
are MMCP candidates. These distinguish themselves by the possibility
of having much larger charges than other MMCPs.

In the following discussion of supersymmetric MMCPs we will tacitly
assume that R-parity is conserved for the scenarios considered unless
stated otherwise. This conserved quantum number, corresponding to
point (i) above, ensures that sparticles can only decay into other
lighter sparticles, resulting in a completely stable Lightest
Supersymmetric Particle (LSP), which as a possible Dark Matter
candidate needs to be neutral. This also prevents the rapid decay of
an MMCP candidate into SM particles.

We can broadly classify all MMCPs on the basis of their electric and
colour charges. This distinction will be important for the discussion
of collider phenomenology in Section~\ref{sec:exp}. For the $N=1$
supersymmetric spectrum we find sleptons and charginos in the first
class, while the colour charged possibilities are gluinos and
squarks. After production, colour charged MMCPs immediately hadronize
into so-called R-hadrons with additional quarks and/or gluons, and may
or may not acquire an electrical charge in the process. Several models
exist in the literature that attempt to describe this hadronization
process.\cite{Drees:1990yw,Baer:1998pg,Gates:1999ei,Mafi:1999dg,Kraan:2004tz,Kilian:2004uj,Kang:2007ib}
For a colour triplet MMCP the possible composition of the R-hadron is
that of a meson- or baryon-like state, {\it e.g.\ }the neutral R-meson
$\tilde t_1\bar d$ or the positively charged R-baryon $\tilde t_1 dd$,
while a colour octet may also form a glueball-like state with a gluon,
$\tilde gg$. R-meson production is dominant in current hadronization
models, and the equivalence of up- and down-quarks means that roughly
half should be charged.

An R-hadron is expected to be surrounded by a small amount of hadronic
energy from the hadronization process and QCD radiation from the
massive particle, forming a jet-structure in the direction of the
MMCP. However, the energy of the jet is limited in magnitude due to
the large mass of the MMCP, and the R-hadron is expected to carry more
than 80\% of the jet energy in the vast majority of
events.\cite{Fairbairn:2006gg}

Let us now briefly discuss the individual MMCP candidates in
supersymmetry that have recieved the most attention in the
literature. The chargino has mostly been considered in scenarios with
Anomaly Mediated Supersymmetry Breaking (AMSB). Here the gaugino
masses have a strict hierarchy $|M_2|\ll |M_1|\ll |M_3|$, so that the
lightest neutralino, which is the LSP, has a dominant wino component,
and the lightest chargino, also dominantly wino, is only marginally
heavier. The mass difference is in the sub-GeV range, but due to model
constraints it is typically above the pion mass for natural parameter
values.\cite{Feng:1999fu,Gherghetta:1999sw}\footnote{The mass
difference in the pure wino case can be though of as the difference
between the charged and neutral wino self-energies, which should be
${\mathcal O}(\alpha m_W)$.}  This means that the decay
$\tilde\chi_1^\pm\to\tilde\chi_1^0\pi^\pm$ is very important to the
collider signature. The chargino decay length $c\tau$ is relatively
restricted and lies between 7~cm at 160~MeV mass difference and 1.6~cm
at 230~MeV.\cite{Chen:1999yf} For a phenomenological study of a
minimal AMSB scenario at the LHC see Ref.~\refcite{Barr:2002ex}.

It is of course possible to ignore the supersymmetry breaking
motivation and resulting constraints of AMSB in the generic Minimal
Supersymmetric Standard Model (MSSM), setting $|M_2|\lsim |M_1|$ and
$|M_2|\ll |\mu|$ by hand at the weak scale to provide a wino LSP,
degenerate with the chargino. In this case the mass difference can
become even smaller, depending on the parameter tuning, allowing for
substantially larger decay lengths dominated by the three-body decay
$\tilde\chi_1^\pm\to\tilde\chi_1^0e^\pm\nu_e$. These are easily
hundreds of meters when the mass difference is less than the pion
mass, making the chargino appear stable in the LHC detectors. Another
option for a chargino MMCP is to have $M_{1,2}\gg |\mu|$, giving a
degenerate higgsino LSP and chargino, which can even be realised in
the focus-point region of the Constrained Minimal Supersymmetric
Standard Model (CMSSM).  When we later discuss the LHC discovery
potential for a chargino MMCP, we will differentiate between the wino
and higgsino cases.

In scenarios with Gauge Mediated Symmetry Breaking (GMSB) and in
supergravity (SUGRA) inspired models, the gravitino may be the
LSP. Any sfermion Next-to-Lightest Supersymmetric Particle (NLSP) has
no choice but to decay to the gravitino, and may thus naturally be
metastable due to the small effective coupling to the gravitino. For
GMSB the suppression is given by $\sqrt{\left<F\right>}$, the effective
supersymmetry breaking scale, and may be as low as $10-100$~TeV. In
SUGRA models the suppression is set by the reduced Planck mass
$M_P=2.4\cdot 10^{18}$~GeV. This leads to a great range of possible
decay lengths for a sfermion NLSP. In GMSB scenarios this has been
estimated as,\cite{Giudice:1998bp}
\begin{equation}
\label{eq:gmsb}
c\tau_{\tilde f}=10^{-2} \left(\frac{100~{\rm GeV}}{m_{\tilde f}}\right)^5
\left(\frac{\sqrt{\left<F\right>}}{100~{\rm TeV}}\right)^4~{\rm cm},
\end{equation}
which, considering realistic values of $\sqrt{\left<F\right>}$ and
sfermion masses, gives decay lengths from sub-micron to
multi-kilometre. The main upper bounds on $\sqrt{\left<F\right>}$ are
of cosmological origin. To avoid ruining Big Bang Nucleosynthesis
(BBN) the sfermion lifetime should be less than $10^4$~s. This implies
that $\sqrt{\left<F\right>}\lsim 10^9$~GeV. For a recent evaluation of
the bounds see Ref.~\refcite{Steffen:2006hw}. For Planck scale
supersymmetry breaking one arrives at a similar result with the
substitution $\left<F\right>\to\sqrt{3}m_{3/2}M_P$ in
Eq.~(\ref{eq:gmsb}).

Third generation sfermions are usually lighter than the two first due
to RGE running effects from a common mass at some high GUT scale, and
as a consequence tend to work better as MMCP candidates in models with
GUT unification of masses. For a given sfermion flavour $\tilde f$, it
is similarly the lightest of the states, $\tilde f_1$, generally a
mixing of the left- and right-handed sfermion, that works as an
MMCP.
However, given mass degeneracy between the NLSP candidates, several
particles may become metastable, and models that are less restricted
at GUT scale may provide different flavors for the NLSP. Still, this
means that it is the lightest stop, sbottom and stau that are most
commonly considered as MMCPs in the literature. In Gravitino Dark
Matter (GDM) scenarios the stau is typically the NLSP for small $m_0$,
giving $m_{\tilde\tau_1}<m_{\tilde\chi_1^0}$, a region otherwise
disallowed due to a charged LSP. If, in addition to small $m_0$, we
take small $\tan(\beta)$ and large $|A_0|$, the NLSP may be the stop,
while large $\tan(\beta)$ and $|A_b|\gg |A_t|$, gives small sbottom
masses.

The NLSP mass degeneracy to a neutralino, or even sneutrino, LSP may
also provide MMCP candidates. However, the needed mass parameters tend
to be fairly fine tuned and unmotivated. Some motivation can be found
in a mass degeneracy that is simultaneously used to explain the
measured Dark Matter density through co-annihilation of LSP and
NLSP. Unfortunately the small mass differences needed for
metastability tend to give too large co-annihilation cross sections
and consequently too little Dark Matter. For a stop NLSP the situation
is a little better because the lifetime of the main decay modes
$\tilde t_1\to c\tilde\chi_1^0$ and $\tilde t_1\to b
ff'\tilde\chi_1^0$ are suppressed by loop-factors and a four-body
phase space, respectively.


The final MMCP candidate we will discuss here is the gluino. In GDM
scenarios the gluino can clearly be metastable like all other
sparticles, if for some reason $M_3$ is small compared to the other
supersymmetry breaking masses so that the gluino is the NLSP, but
perhaps more interesting is the split-SUSY scenario where the scalars
--- except the lightest Higgs that decouples --- are much heavier than
the gauginos. This ignores the Higgs mass hierarchy problem, but
naturally suppresses CP violation and avoids constraints from flavor
physics.  Since the gluino can only decay through virtual squarks,
either to two quarks and a gaugino, or in a loop-decay to a gluon and
neutralino, the decay is suppressed by the mass ratio $m_{\tilde
g}^2/m_{\tilde q}^2$, giving the scaling
behaviour,\cite{Toharia:2005gm,Gambino:2005eh}
\begin{equation}
c\tau_{\tilde g}=10^{11} \left(\frac{1~{\rm TeV}}{m_{\tilde g}}\right)^5
\left(\frac{m_{\tilde q}}{10^9~{\rm GeV}}\right)^4~{\rm cm}.
\end{equation}
The CMSSM focus-point region with sufficiently large values for $m_0$
displays these properties, but the constraint on electro-weak symmetry
breaking means that $m_{1/2}$ must also be rather large, and the
sparticle mass spectrum might be outside of the reach of the LHC.

With non-zero R-parity violating couplings the LSP will eventually
decay unless it is very light. If the LSP is the gravitino it may
still live long enough to constitute Dark
Matter.\cite{Takayama:2000uz,Buchmuller:2007ui,Lola:2007rw} For
couplings less than ${\mathcal O}(10^{-6})$, the exact value depending
on sparticle mass and flavor of the coupling, the lifetime of stable
or metastable sparticles is not shortened enough for them to decay
inside a detector, and thus has no direct influence on the collider
phenomenology of MMCPs. However, a new and interesting possibility
opens up: the LSP may well be charged if it has a short lifetime on
cosmological scales. This allows for less fine-tuning for many MMCP
candidates discussed above that formerly decayed to the LSP, they now
simply need to be the lightest sparticle.

\section{Experimental Issues}
\label{sec:exp}

The signatures of MMCPs traversing a detector are determined by their
interactions with the matter of the detector. The main source of
energy loss for electrically charged massive particles with large
velocities, $\beta\gamma\gsim 0.1$, is ionisation energy loss through
the removal of electrons from atoms in the material. For a particle of
velocity $\beta\gamma$ this is given by the Bethe-Bloch formula for
mean rate of ionisation energy loss, or stopping
power,\cite{Amsler:2008zzb}
\begin{equation}
\left<-\frac{dE}{dx}\right>=\frac{Kz^2Z}{A\beta^2}\left(\frac{1}{2}
\ln\frac{2m_ec^2\beta^2\gamma^2T_{\max}}{I^2}-\beta^2-\frac{\delta}{2}\right), 
\label{eq:BB}
\end{equation}
where $K=4\pi N_Ar_e^2m_ec^2$. The physical constants and properties
of the material and particle involved are: the electric charge $z$ of
the particle, the atomic number $Z$ and mass $A$ of the material, the
classical electron radius $r_e$ and mass $m_e$, Avogadro's number
$N_A$, the maximum kinetic energy that can be given to a free electron
in a single collision $T_{\max}$, and the mean ionisation potential of
the material $I$.\footnote{Note that $x$ takes the form of length
times density of the material and the units are given by
$K/A=0.307075$~MeV~g$^{-1}$~cm$^2$ for an atomic mass of
$A=1$~g~mol$^{-1}$.}

The $\delta$ term in Eq.~(\ref{eq:BB}) is a function of velocity that
corrects for density effects at relativistic velocities,
$\beta\gamma\gsim 3$. In addition to the ionisation energy loss,
lighter charged particles also have significant radiative losses at
high velocities. For muons radiative effects reach 1\% at around
10~GeV energy. However, this is mass dependent, and should be
insignificant for masses not already ruled out for MMCPs.

From Eq.~(\ref{eq:BB}) one finds a minimum around $\beta\gamma\simeq
3$, roughly independent of material and particle mass, for masses much
greater than the electron mass. Such a particle is usually termed a
minimum ionising particle or mip. The asymptotic behaviour
$dE/dx\propto\beta^{-2}$ of the Bethe-Bloch formula at low velocities
is tempered for velocities less than the bound electron velocity, of
the order of a few percent of $c$, where the energy loss is simply
proportional to $\beta$.

The fact that the ionisation energy loss of a charged particle is
almost independent of the particle's mass, but dependent on its
velocity is crucial in searches for MMCPs at colliders. Firstly, it
means that the MMCPs will appear as muon-like objects in detectors
and their momenta can be reconstructed by a tracking system that
relies on ionisation energy deposits. Secondly, while the energy loss
of an MMCP produced may very well be that of a mip, an accompanying
measurement of the particle's momentum $p$ from the curvature of its
track in a magnetic field will reveal that the mass $m$ is large since
$m=p/\beta\gamma$, and as we have seen above, a measurement of $dE/dx$
fixes the velocity modulo fluctuations in single measurements.

For strong interactions of MMCPs with detector material the situation
is much less well understood as we lack sufficiently similar SM
counterparts. Several models have been proposed that attempt to
describe the scattering of R-hadrons on atomic nuclei.
\cite{Drees:1990yw,Baer:1998pg,Gates:1999ei,Mafi:1999dg,Kraan:2004tz,Kilian:2004uj,Kang:2007ib,Mackeprang:2006gx,deBoer:2007ii} The common feature of these models is the spectator
nature of the MMCP. Its interactions with partons in the nuclei are
suppressed relative to those of the light constituents of the R-hadron
by the square of the MMCP mass. The implication is that an R-hadron's
properties are largely determined by its light-quark/gluon components.

Despite the spectator role of the MMCP it carries the majority of the
R-hadron momentum since the relative momentum of the interacting
system is given by the ratio of its mass to the total R-hadron
mass. Thus, for an MMCP mass of ${\mathcal O}(100~{\rm GeV})$ or
above, the momentum of the scattering system is maximally a few GeV at
the LHC. The low-energy scatters means that the energy loss per
interaction is also limited, of the order of 1~GeV over a large
spectrum of $\beta$. While the energy loss to strong interactions is
still expected to dominate at high velocities, the rise in ionisation
energy loss below the mip velocity means that it dominates at low
velocities.

One important consequence of the limited energy loss is that most
strongly interacting MMCPs are expected to penetrate large detectors,
their tracks, if charged, being in effect reconstructed as muons. The
additional complication with R-hadrons is that they may flip charge in
inelastic scattering off nuclei, interchanging their non-MMCP partons
with those of the nuclei. Thus R-hadron tracks in detectors can be
stubs that end or appear abruptly in the tracking system. Recently it
has also been shown that R-mesons will tend to convert into R-baryons
due to scatterings of the type $R_{\tilde gq\bar q}+ N\to R_{\tilde
gqqq}+\pi$ that are energetically favoured, and where the reverse
reaction is suppressed due to a lack of pions in the detector
material.\cite{Kraan:2004tz}

The second staple of MMCP searches is the Time-of-Flight (ToF)
measurement. With the use of high magnetic fields it is safe to assume
that any stable or metastable SM particles reconstructed by a tracking
system travel with a velocity of $\beta\simeq 1$. Despite this, due to
the large size of modern detector experiments, the ToF from the
interaction point and out of the detector can be of ${\mathcal
O}(100~{\rm ns})$. Given a time resolution in the nanosecond range,
particle velocities can then be reconstructed with good
precision. Since massive particles are produced with much smaller
velocities, these can be discriminated from SM backgrounds, and again
the combination of momentum and velocity measurements leads to a
measurement of the mass.

While the low velocity of the MMCPs is an advantage for measuring
masses, it poses serious challenges for triggering on signal events
and the acquisition of data in modern high luminosity colliders. This
is particularly true for the LHC, where the distance between two bunch
crossings (BCs) is only $25$~ns at design luminosity. This corresponds
to $7.5$~m at light-speed, meaning that up to three BCs are
simultaneously contained inside the large ATLAS and CMS
detectors. This is taken into account in detector design and read-out,
however, assuming particle velocities near $c$. Slower particles may
cause triggers to be assigned to the wrong BC, or the MMCP not to be
recorded in the triggered BC.\cite{Tarem1} Simple estimates from ATLAS
geometry give that MMCPs with $\beta>0.7$ reach the outer parts of the
muon trigger system inside the correct BC.\cite{ERO2} This is
confirmed in a full simulation study of MMCPs that show large
efficiencies for hits in the muon trigger chambers at $\beta\gsim
0.7-0.8$ with standard trigger windows.\cite{Tarem:2009zz,Aad:2009wy}
If MMCPs are indeed discovered at the LHC, trigger windows can be
enlarged to values used in calibration and debugging to improve
efficiencies and reach lower velocities.

For the particular case of R-parity conserving supersymmetric models,
MMCPs are produced in pairs at colliders, either through cascade
decays or direct production. This means that there are at least two
possible muon trigger objects in each event, which improves the final
trigger efficiency, in particular for cascade decays where the MMCPs
may be produced with wildly different velocities. The exception is
the R-hadron scenario, where one or both of the R-hadrons may be
neutral by the time they pass through the muon system, either from
being produced neutral or by charge flipping interactions.

The second difficulty with slow moving MMCPs is the drop in track
reconstruction efficiency. Tracking software also operates under the
$\beta=1$ assumption, so that slow particles typically have a bad
quality of fit for their tracks and may be discarded, {\it e.g.\ }for
the Monitored Drift Tubes (MDTs) in the ATLAS muon system, hits from
slower particles are assigned too large drift distances from the tube
centre so that the reconstruction is attempted with mismeasured
coordinates.  As a result the reconstruction efficiency is found to
decrease dramatically below
$\beta=0.75$.\cite{ERO2,Tarem:2009zz,Aad:2009wy} The solution
suggested in Ref.~\refcite{Polesello} is to keep $\beta$ a free
parameter of the fit and maximise the fit quality over $\beta$. This
simultaneously allows for better reconstruction efficiencies and a
determination of the MMCP velocity. In a full simulation study, using
also information from the timing of hits in the muon trigger system
and recovering trigger hits from the next BC,
Ref.~\refcite{Tarem:2009zz} finds good reconstruction efficiencies
down to $\beta\simeq 0.5$.

\section{Current Constraints}
\label{sec:bounds}

The best current collider bounds on MMCP cross sections and masses
originate from the OPAL experiment at LEP and recent results published
by the Tevatron CDF and D0
experiments.\cite{Abbiendi:2003yd,Aaltonen:2009ke,Abazov:2008qu} The
lower mass bounds are typically the result of a comparison between the
predicted pair-production cross section of an MMCP in a particular
model, and a limit on the number of events with particles
reconstructed above a certain mass from the experiment. Below we
review these bounds and comment briefly on the assumptions going into
them. In general, the reliance on pair-production cross sections gives
conservative mass bounds for sparticle MMCP candidates, with little or
no model dependence. If interpreted in specific constrained models,
the Tevatron bounds can improve dramatically, in particular for weakly
interacting sparticles, where the dominant production is expected to
be from the decays of strongly interacting sparticles with much higher
cross sections.

The OPAL experiment uses a measurement of $dE/dx$ for each track
compared to the momentum to distinguish massive charged particles from
SM backgrounds. The ionisation energy loss is measured with a good
resolution of 2.8\%, using multiple, above 20 required, hits on sense
wires in the OPAL jet chamber. The results are based on data collected
at centre-of-mass energies from 130~GeV to 209~GeV, with a total
integrated luminosity of $693.1$~pb$^{-1}$. In the absence of any
signal, a 95\% C.L.\ bound on the pair-production cross section of
weakly interacting MMCPs is given in terms of the MMCP mass and spin,
assuming $s$--channel production. Translated into sparticle masses
this implies a lower limit of 98.0~GeV and 98.5~GeV on right- and
left-handed sleptons, respectively, for smuons and staus. Note that
this bound does not translate to selectrons, as their cross section
depends on other sparticle parameters in $t$--channel
production. The same model-dependence is also present for chargino
pair-production in terms of the wino--higgsino mixing and the
sneutrino mass. By minimising the chargino cross section over the
CMSSM parameter space for a fixed chargino mass, a lower limit of
102.0~GeV is found.


For squark MMCPs the ALEPH experiment at LEP has given direct mass
bounds, with the conservative assumption that stop and sbottom mixing
angles are such that their couplings to the $Z$
disappear.\cite{Heister:2003hc} The resulting lower limits on the
lightest mass eigenstates are $m_{\tilde t_1}>95$~GeV and $m_{\tilde
b_1}>92$~GeV at 95\% C.L. For a gluino MMCP, the bounds from LEP are
much weaker. The dominant production process is $e^+e^-\to q\bar
q\tilde g\tilde g$, where a gluon radiating off one of the quarks
splits into two gluinos. The ALEPH collaboration gives a bound of
$m_{\tilde g}>26.9$~GeV.\cite{Heister:2003hc}

The Tevatron CDF experiment reports complementary bounds for weakly
interacting MMCPs and more stringent bounds for strongly interacting
MMCPs set with $1$~fb$^{-1}$ of data at $\sqrt{s}=1.96$~TeV. The CDF
experiment uses a ToF detector to measure the masses of particles with
high transverse momentum in a sample of events collected with the muon
trigger. Standard Model backgrounds are estimated from the mass
resolution on muon-like particles found in a control-region, and no
excess of events is found for masses above $100$~GeV. Along with an
evaluation of the signal efficiency, this allows fully model
independent cross section bounds of $\sigma<10$~fb and $\sigma<48$~fb
to be set on weakly and strongly interacting MMCPs, respectively, for
a single MMCP within the experimental cuts of $p_T>40$~GeV,
$|\eta|<0.7$, $0.4<\beta<0.9$ and $m>100$~GeV. Note that the CDF
analysis does not consider the R-meson to R-baryon transition
mentioned in Section~\ref{sec:exp} when evaluating the signal
efficiency. The R-baryon should have a larger interaction cross
section, leading to lower efficiency, so that the final mass limit may
be slightly overestimated.

For an interpretation of these bounds in terms of sparticle masses,
the differential pair-production cross sections need to be evaluated
and integrated over the acceptance region. This was done for a stop
MMCP candidate, yielding a lower bound of $m_{\tilde t_1}>249$~GeV at
the 95\% C.L.\ for a specific supergravity inspired scenario where the
other squark masses are set to $m_{\tilde q}=256$~GeV, the gluino mass
$m_{\tilde g}=284$~GeV and the stop mixing angle
$\sin(2\theta)=-0.99$. However, the stop pair-production cross section
at NLO is known to be rather insensitive to these parameters, see
Ref.~\refcite{Beenakker:1997ut}. We evaluate this effect using {\sc
Prospino
2.1},\cite{Beenakker:1997ut,Beenakker:1996ch,Beenakker:1999xh}
decoupling the contribution from gluinos and other squarks by setting
their masses to 2~TeV and minimising the cross section over the mixing
angle. As expected, this gives only a slightly softer bound of
$m_{\tilde t_1}>241.5$~GeV.

No explicit mass limits for other supersymmetric MMCPs were found in
Ref.~\refcite{Aaltonen:2009ke}. For the gluino in particular the CDF
data should result in an improvement of the LEP mass limit, and its
evaluation would be feasible with Monte Carlo simulation of the
kinematics of gluino pair-production and a re-evaluation of the signal
efficiency for a gluino based R-hadron. Similar considerations apply
to the sbottom, where the cross section should be identical to the
stop's in the decoupling limit, but where the signal efficiency for an
R-hadron containing a down-squark could be different due to the
preferred formation of a neutral R-baryon.

The D0 experiment presents a similar analysis using $1.1$~fb$^{-1}$ of
data, measuring particle velocity through ToF with timing from
scintillation trigger counters in the muon system. The results are
less model independent in that they are interpreted directly in terms
of upper cross section bounds on the pair-production of staus and
charginos. For the stau, no mass limit can be set as the predicted
cross section is too low. For the chargino, mass limits of 206~GeV and
171~GeV, for a wino- and higgsino-like chargino, respectively, is
given at the 95\% C.L.

We give a summary of the best limits discussed above in
Table~\ref{tab:bounds}. In addition to these bounds there are also
indirect bounds such as the measurement of the invisible $Z$ width,
which constrains the mass of supersymmetric MMCPs that couple to the
$Z$ to be above $m_Z/2$,\cite{Amsler:2008zzb} and the running of
$\alpha_s$, which limits colour charged particles beyond the SM to
have masses greater than about 6~GeV.\cite{Csikor:1996vz} While these
limits are less constraining than the direct bounds discussed above,
they have the advantage of closing low-mass loopholes in the direct
bounds.
As mentioned in Section~\ref{sec:models} there are also limits on the
MMCP lifetime from cosmology that translate to mass limits, given
model dependent assumptions on couplings and/or other sparticle
masses.

\begin{table}[h]
\tbl{Summary of current best limits on MMCP masses in supersymmetry.}
{\begin{tabular}{@{}ccll@{}}
\toprule
MMCP & Mass limit [GeV] & Notes & Source\\
\colrule
$\tilde\chi_1^\pm$ & 206.0 & Assuming $\tilde\chi^\pm_1\simeq\tilde W^\pm$.
                   & D0, Ref.~\refcite{Abazov:2008qu} \\
                   & 171.0 & Assuming $\tilde\chi^\pm_1\simeq\tilde H^\pm$. 
                   & D0, Ref.~\refcite{Abazov:2008qu} \\
$\tilde\ell_1$     &  98.0 & $\tilde\ell=\tilde\mu,\tilde\tau$, assuming
                   $\tilde\ell_1\simeq\tilde\ell_R$.
                   & OPAL, Ref.~\refcite{Abbiendi:2003yd} \\
                   &  98.5 & $\tilde\ell=\tilde\mu,\tilde\tau$, assuming
                   $\tilde\ell_1\simeq\tilde\ell_L$.
                   & OPAL, Ref.~\refcite{Abbiendi:2003yd} \\
\colrule
$\tilde g$         &  26.9 & Could be improved using Tevatron data.
                   & ALEPH, Ref.~\refcite{Heister:2003hc} \\
$\tilde t_1$       & 241.5 & Re-evaluated in this review. 
                   & CDF, Ref.~\refcite{Aaltonen:2009ke} \\
$\tilde b_1$       & 92.0 & Could be improved using Tevatron data.
                   & ALEPH, Ref.~\refcite{Heister:2003hc} \\
\botrule
\label{tab:bounds}
\end{tabular}}
\end{table}

\section{LHC Prospects}
\label{sec:LHC}

With the large increase in centre-of-mass energy from the Tevatron to
the LHC, the LHC experiments will naturally be sensitive to more
massive MMCPs. We begin this Section with a discussion of the
prospects for applying $dE/dx$ and ToF techniques to discovering MMCPs
and measuring their masses in the large LHC experiments. We then show
an estimate of the discovery reach of the LHC in terms of sparticle
MMCP masses, before we touch on the exciting possibility of stopped
MMCPs.


In both ATLAS and CMS the principal sub-detectors for measuring
$dE/dx$ are in the inner tracking systems. Although more energy will
deposited in the calorimeters, a reliable determination of $dE/dx$
needs several measurements on account of large fluctuations.
In published ATLAS results the current optimal strategy uses a
measurement of the time a hit in the Transition Radiation Tracker
(TRT) is over an ionisation threshold.\cite{Kraan:2005ji} This is
strongly correlated with $dE/dx$ and can be used on slow particles,
$\beta\lsim 0.7$, where from the Bethe-Bloch formula we have
$dE/dx\propto
\beta^{-2}$. Results given in Ref.~\refcite{Driouichi} indicate that
in this region a resolution on the velocity of somewhat better than
20\% is achievable. For CMS, the inner tracking system consists of
silicon pixel and silicon strip detectors, the $dE/dx$ resolution of
these devices is somewhat worse than for gaseous detectors such as the
ATLAS TRT.

Assuming that the error on the momentum and velocity measurements are
uncorrelated, the mass resolution is given by
\begin{equation}
\left(\frac{\sigma_M}{M}\right)^2=\left(\frac{\sigma_p}{p}\right)^2
+\left(\gamma^2\frac{\sigma_\beta}{\beta}\right)^2.
\label{eq:mass_res}
\end{equation}
As expected the resolution becomes very bad for $\beta\to 1$. With a
typical transverse momentum resolution of $4-10\%$ for the 100~GeV to
1~TeV momentum range in ATLAS,\cite{:2008zzm} we see that the mass
resolution is dominated by the $\beta$--resolution. Thus a 20\%
resolution on the MMCP mass should realistically be achievable in
ATLAS using the time-over-threshold technique.


For a mass measurement using ToF, the error in the velocity
measurement is dominated by the timing resolution since the positions
of hits inside a detector will be very well known from an extensive
calibration programme. Thus we can substitute
$\sigma_\beta/\beta=\sigma_t/t$ in Eq.~\ref{eq:mass_res}. From a
simple order of magnitude estimate, using the size of a generic LHC
detector and a $1$~ns fluctuation on the time measurement, we get a
resolution of $\sigma_t/t\simeq 0.02$, showing the advantage of the
size of the LHC detectors. This matches well with various full
simulation studies of the ATLAS detector using the track re-fit method
described in Section~\ref{sec:exp}, timing information from the muon
trigger system, or
both.\cite{ERO2,Tarem:2009zz,Aad:2009wy,Polesello,Marianne} These
arrive at similar resolutions of $\sigma_\beta/\beta=0.02-0.03$ for a
stau MMCP, which is competitive with the contribution to the mass
resolution from the momentum measurement at low velocities
($\beta\lsim 0.8$).

The final precision with which the MMCP mass can be determined is of
course to some degree dependent on the available statistics, the MMCP
mass, and to an extent on whether the MMCP has electric and/or colour
charge.\footnote{Because of interactions with nuclei, R-hadrons lose
more energy in the calorimeters. Charge flipping also makes precise
momentum and velocity determination more difficult.} However, the good
resolution achievable with ToF techniques quickly reduces the
statistical error to below 1~GeV, so that the mass determination is
dominated by systematic errors that have been found to be around
$1-2$~GeV in Monte Carlo simulations.\cite{ERO2,Tarem:2009zz} This
level of precision has been shown to allow the efficient
reconstruction of many heavier sparticles, if not too heavy to be
produced, that decay into the MMCP, giving precise access to a large
fraction of the sparticle spectrum.\cite{Ellis:2006vu}


At present little has published by the LHC experiments on their
potential discovery reach in terms of MMCP masses; exceptions are
found in Refs.~\refcite{Aad:2009wy}, \refcite{Kraan:2005ji} and
\refcite{CMS1}. This is understandable as potential main backgrounds
such as muons with mis-reconstructed velocities are difficult to
estimate from Monte Carlo alone, without the benefit of the detector
understanding that comes with data.  Nevertheless, we can attempt a
rough estimate the LHC reach under the assumption that an MMCP search
at the LHC will be a low background search, with ${\mathcal O}(1)$
background events expected for some fixed integrated luminosity and a
corresponding signal efficiency; this is based on similar analysis at
the Tevatron and at LEP, described in Section~\ref{sec:bounds}.

\begin{figure}[h]
\centerline{\psfig{file=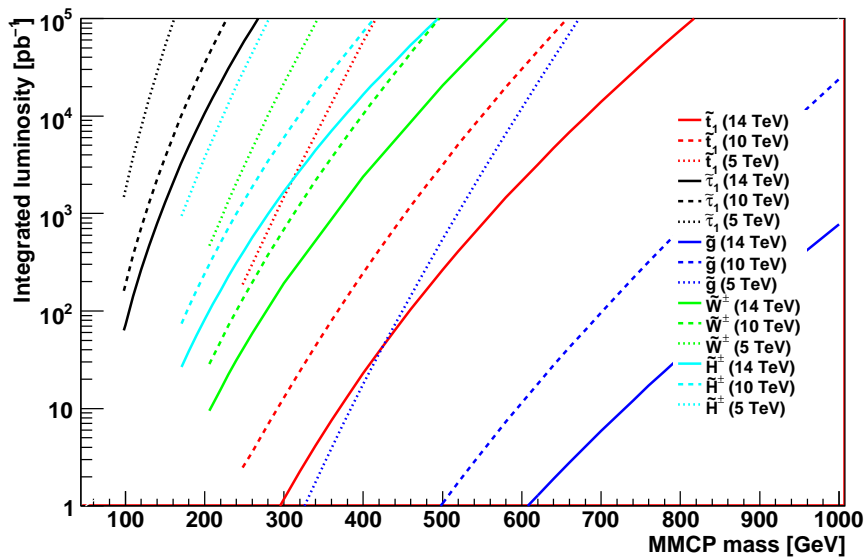,width=\linewidth}}
\vspace*{6pt}
\caption{Integrated luminosities needed for discovery versus MMCP mass
for the LHC at 14~TeV (solid), 10~TeV (dashed) and 5~TeV (dotted). The 
colours denote different sparticle species: stop (red), stau (black), 
gluino (blue), charged wino (green) and higgsino (cyan).
\protect\label{fig:reach}}
\end{figure}

We adopt a fairly conservative guess for the signal efficiency of 20\%
and 5\%, for electrically charged and strongly interacting MMCPs,
respectively, based on CMS numbers for a search with first
data.\cite{CMS1} With the NLO pair-production cross sections computed
by {\sc Prospino 2.1} in the limit where other sparticles effectively
decouple, we show the integrated luminosity needed for ``discovery''
as a function of MMCP mass in Fig.~\ref{fig:reach}. The required
luminosity is of course dependent on the expected number of background
events for a given luminosity, and we have assumed $B=1$ for $100~{\rm
pb}^{-1}$, but the required luminosity scales linearly with $B$ if one
takes $S/\sqrt{B}$ as the significance. In the same way the interested
reader can easily re-scale Fig.~\ref{fig:reach} to a different signal
efficiency, since the required integrated luminosity scales as
$\epsilon^{-2}$.

While the above discussion has focused mostly on properties of the
ATLAS detector, the situation for CMS is similar, with somewhat better
signal efficiencies expected due to its ``compact'' design, and a
somewhat worse mass resolution for the same
reason.\cite{Zalewski,CMS1,Chen} For the smaller experiments work on
MMCPs has been limited, but it is possible that the LHCb detector may
be competitive in some areas of parameter space due to excellent
particle identification properties. The LHCb will certainly be in an
good position to observe long-lived particles decaying inside the
detector, a topic which is outside the scope of this paper. In
addition, the MOEDAL experiment has proposed to search for highly
ionising exotic particles such as magnetic monopoles and Q-balls,
through the use of a passive plastic track-etch detector surrounding
the LHCb vertex detector.\cite{Pinfold:1999sp}

Finally in this Section, we want to mention a very interesting
possibility due to MMCP energy loss: MMCPs produced with small momenta
may lose enough energy to ionisation to be stopped inside the
detector, most probably inside a calorimeter, or in a dedicated
stopper detector as suggested in Refs.~\refcite{Feng:2004yi},
\refcite{DeRoeck:2005bw} and \refcite{Hamaguchi:2006vu}. Due to the small
energy losses expected the fraction of stopped MMCPs will be a small
unless the MMCP is very heavy, in which case the small cross section
limits the number of stopped MMCPs. Even if rare, stopped MMCPs may
open up a very exciting new direction of exploration: that of
detecting late decays of MMCPs, with lifetimes from a fraction of a
second to years. This idea has already seen initial exploration at the
Tevatron D0 experiment, looking for stopped
gluinos.\cite{Arvanitaki:2005nq,Abazov:2007ht}

Decays in the detector out of sync with collisions, or even with no
beam present, will enable a measurement of the MMCP lifetime and its
dominant decay channel(s), if the decays can be separated from cavern
backgrounds and cosmic ray hits, and enough statistics is
available. In models with a gravitino LSP the gravitino mass could
also be measured given knowledge of the MMCP mass and the recoil
energy against an invisible gravitino in the MMCP decay. The
implication of all this is that the coupling of the MMCP to the
gravitino will be indirectly determined from the MMCP lifetime, which
constitutes a microscopic measurement of Newton's constant, and a
powerful piece of evidence in favour of
supergravity.\cite{Buchmuller:2004rq}


\section{Conclusion}
\label{sec:conclusion}
We have given a brief review of the prospects for discovering Massive
Metastable Charged Particles at the LHC, focusing on supersymmetric
scenarios. The unprecedented centre-of-mass energy of the LHC allows
mass scales far above current constraints to be probed even with
limited statistics. The discovery of such a particle will herald a
revolution in particle physics: measuring the properties of the MMCP
will give clear indications of the structure of physics beyond the
Standard Model, in the same way that the properties of stable and
metastable Standard Model particles have helped reveal the structure
of that model.

The general purpose LHC detectors are well suited to measure the mass
and charge of an MMCP as a result of their size and the excellent
timing and momentum resolution in several sub-detectors. If stopped in
the detectors, the MMCP lifetime can also be determined, with possible
implications even for Planck scale physics.

\section*{Acknowledgments}

ARR acknowledges funding from the UK Science and Technology Facilities
Council (STFC) and wishes to thank John R.\ Ellis and Ola K.\ \O ye
for fruitful collaborations on this topic, as well as David A.\
Milstead and members of the Cambridge Supersymmetry Working Group for
many interesting and helpful discussions.

\end{document}